\documentclass[~,twocolumn,amsmath,amssymb,reprint,nofootinbib]{revtex4-2}
\usepackage{epsfig}                                                                 % Include eps figure files
\usepackage{dcolumn}                                                                % Align table columns on decimal point
\usepackage{bm}                                                                     % Bold math

\usepackage{placeins}

\PassOptionsToPackage{hyphens}{url}
\usepackage[pdfstartview=FitH,bookmarksopen=true,bookmarksopenlevel=2]{hyperref}    % Include hypertext links with bookmarks
\usepackage{url}

\begin{document}
\title{Supercooling of liquids, as described by the Enskog--Vlasov kinetic equation}
\author{E. S. Benilov}
 \email[Email address: ]{Eugene.Benilov@ul.ie}
 \homepage[\newline Homepage: ]{https://eugene.benilov.com/}
 \affiliation{Department of Mathematics and Statistics, University of Limerick, Limerick V94~T9PX, Ireland}

\begin{abstract}
A model combining Enskog's collision integral for dense fluids with a
Vlasov-style description of the van der Waals force is applied to
supercooling. First, the spinodal temperature $T_{s}$ is calculated, at which
a liquid becomes unstable to small perturbations and transitions to solid. In particular, it
turns out that isochoric cooling allows one to reach a lower temperature than
isobaric cooling. Second, the surface tension of a supercooled liquid--vapor
interface is shown to diverge at $T_{s}$. The singularity is caused by an
oscillatory region emerging on the liquid side of the interface as
$T\rightarrow T_{s}$; it develops because the liquid approaches instability,
and the interface starts radiating (so far, evanescent) waves. At $T=T_{s}$,
the waves cease to be evanescent and the oscillatory region extends to
infinity---hence, the singularity of the surface tension. Since this effect
has a clear physical interpretation, it should occur regardless of the model
and approximations under which it was obtained. This and the other results of
the paper are illustrated using argon and several other fluids.

\end{abstract}
\maketitle

\section{Introduction\label{sec1}}

The Enskog--Vlasov (EV) kinetic equation was proposed in 1971 by Grmela
\cite{Grmela71} as a tool for studying phase transitions. It differs from the
classical Boltzmann equation \cite{Boltzmann70} in two respects. First, it
includes a term describing the van der Waals force, formulated in the spirit
of Vlasov's description of electromagnetic forces in plasmas \cite{Vlasov68}.
Second, it employs Enskog's collision integral \cite{Enskog22} for
\emph{dense} fluids, instead of Boltzmann's integral for \emph{dilute} fluids.
Since 1971, the EV equation has been used in both theoretical (e.g., Refs.
\cite{BenilovBenilov18,BenilovBenilov19a,BenilovBenilov19b,StruchtrupFrezzotti22}%
) and computational (e.g., Refs.
\cite{BarbanteFrezzottiGibelliLegrenziCoriglianoFrangi12,BarbanteFrezzottiGibelli15,WuZhangReese15,SadrGorji19,BusuiocFrezzottiGibelli23}%
) studies.

Note that the Enskog--Vlasov equation can be derived from first principles
only for a fluid of hard spheres \cite{VanbeijerenErnst73a}. For real fluids,
it should be viewed as a phenomenological model and, thus, needs to be
calibrated---i.e., its parameters should be chosen to ensure that the
properties of the `theoretical' fluid match those of the `real' one. Most
importantly, the EV model can be fully calibrated using the fluid's critical
and triple-point parameters. These \emph{macroscopic} characteristics are well
defined and easier to measure than, say, the \emph{microscopic} free-energy
barrier in the classical nucleation theory. More generally, a theory
calibrated on the very parameters it predicts is vulnerable to the accusation
of being merely a data fit, not reflecting the physics of the phenomenon in question.

In this work, the Enskog--Vlasov equation is used to calculate the parameters
of supercooled fluids. The simplest version of the EV model will be used---the
one for spherical molecules---so that the rotational degrees of freedom can be
ignored. The results are illustrated using argon. To see how deviation from
spherical symmetry affects supercooling, several diatomic fluids are
considered: nitrogen, oxygen, fluorine (all with quadrupolar, axisymmetric
molecules), and carbon dioxide (dipolar, axisymmetric). The nonrotational EV
model describes their thermodynamic properties surprisingly well (see Appendix
\ref{appA} ), but their heat capacities remain inaccurate, and so the results
obtained for these fluids should be viewed as indicative only. It is also
worthwhile---and requires little extra work---to consider one example of a
triatomic fluid, namely water (admittedly, the EV model approximates its
properties less accurately than those of the other fluids considered,
particularly its surface tension). It can, however, be argued that some of the
qualitative results obtained for polyatomic fluids using the nonrotational EV
equation have a clear physical interpretation and should therefore persist in
a more accurate model.

The paper has the following structure. In Sect. \ref{sec2}, the Enskog--Vlasov
model is formulated and calibrated. In Sect. \ref{sec3}, the model is applied
to flat liquid--vapor interfaces, and in Sect. \ref{sec4}, to spinodal
instability (when liquid transitions to solid). The results are compared to
those of experiments and molecular dynamics simulations in Sect. \ref{sec5}.

\section{Formulation\label{sec2}}

\subsection{The Enskog--Vlasov kinetic equation\label{sec2.1}}

Consider a fluid characterized by the distribution function $f(\mathbf{v}%
,\mathbf{r},t)$, where $\mathbf{v}$ is the molecular velocity, $\mathbf{r}$
the position vector, and $t$ the time. Let the fluid's molecules have mass $m$
and be affected by a force $\mathbf{F}(\mathbf{r},t)$.

The generic form of a kinetic equation is%
\begin{multline}
\frac{\partial f(\mathbf{r},\mathbf{v},t)}{\partial t}+\mathbf{v}%
\cdot\mathbf{\nabla}f(\mathbf{r},\mathbf{v},t)\\
+\frac{1}{m}\mathbf{F}(\mathbf{r},t)\cdot\frac{\partial f(\mathbf{r}%
,\mathbf{v},t)}{\partial\mathbf{v}}=\mathrm{St}[f], \label{2.1}%
\end{multline}
where the collision integral $\mathrm{St}[f]$ is a functional of
$f(\mathbf{v},\mathbf{r},t)$.

To specify $\mathrm{St}[f]$, consider two molecules with velocities
$\mathbf{v}$ and $\mathbf{v}_{1}$, which collide and change their velocities
to $\mathbf{v}^{\prime}$ and $\mathbf{v}_{1}^{\prime}$. Approximating the
molecules by hard spheres, one can readily show that%
\begin{equation}
\mathbf{v}^{\prime}=\mathbf{v}+\mathbf{u}\left(  \mathbf{g}\cdot
\mathbf{u}\right)  ,\qquad\mathbf{v}_{1}^{\prime}=\mathbf{v}_{1}%
-\mathbf{u}\left(  \mathbf{g}\cdot\mathbf{u}\right)  , \label{2.2}%
\end{equation}
where $\mathbf{u}$ is the unit vector directed from the centre of the sphere
with velocity $\mathbf{v}_{1}$ toward the impact point, and%
\begin{equation}
\mathbf{g}=\mathbf{v}_{1}-\mathbf{v}. \label{2.3}%
\end{equation}
Introduce also the scattering cross-section $\sigma$ and note that, for hard
spheres of diameter $D$, it is given by%
\begin{equation}
\sigma=D^{2}\frac{\mathbf{g}\cdot\mathbf{u}}{\left\vert \mathbf{g}\right\vert
}\operatorname{H}(\mathbf{g}\cdot\mathbf{u}), \label{2.4}%
\end{equation}
where $\operatorname{H}$ is the Heaviside step function.

In the present paper, the following collision integral is
used:\begin{widetext}%
\begin{equation}
\mathrm{St}=\int\int\left[  \eta(\mathbf{r},\mathbf{r}+D\mathbf{u}%
,t)\,f(\mathbf{r},\mathbf{v}^{\prime},t)\,f(\mathbf{r}+D\mathbf{u}%
,\mathbf{v}_{1}^{\prime},t)-\eta(\mathbf{r},\mathbf{r}-D\mathbf{u}%
,t)\,f(\mathbf{r},\mathbf{v},t)\,f(\mathbf{r}-D\mathbf{u},\mathbf{v}%
_{1},t)\right]  \left\vert \mathbf{g}\right\vert \sigma\,\mathrm{d}%
^{2}\mathbf{u}\,\mathrm{d}^{3}\mathbf{v}_{1},\label{2.5}%
\end{equation}
\end{widetext}where the coefficient $\eta(\mathbf{r},\mathbf{r}_{1},t)$
describes high-density effects. To date, it has been calculated from first
principles only for a fluid of hard spheres \cite{VanbeijerenErnst73a} and is
otherwise treated as an adjustable parameter. As such, it should still comply
with fundamental requirements: the Onsager reciprocal relation, the H theorem,
and the correct dilute-gas limit: $\eta\rightarrow1$ as $D\rightarrow0$ (so
that (\ref{2.5}) reduces to the standard Boltzmann integral). To emphasize the
importance of these requirements, note that the original version of Enskog's
collision integral \cite{Enskog22} was shown \emph{not} to satisfy the Onsager
relations and the H theorem \cite{LebowitzPercusSykes69}. It was subsequently
modified \cite{VanbeijerenErnst73a,VanbeijerenErnst73b}, and the new version
does satisfy both \cite{VanbeijerenErnst73a,VanbeijerenErnst73b,Resibois78}
(it is actually the one which describes the fluid of hard spheres).

In this paper, the following expression for $\eta$ is used:
\begin{widetext}%
\begin{equation}
\eta(\mathbf{r},\mathbf{r}_{1},t)=1+\sum_{l=2}^{L}c_{l}\int^{l-1}\left[
%TCIMACRO{\dprod _{i=2}^{l}}%
%BeginExpansion
{\displaystyle\prod_{i=2}^{l}}
%EndExpansion
n(\mathbf{r}_{i},t)\,\operatorname{H}(D-\left\vert \mathbf{r}%
-\mathbf{\mathbf{r}}_{i}\right\vert )\,\operatorname{H}(D-\left\vert
\mathbf{r}_{1}-\mathbf{\mathbf{r}}_{i}\right\vert )\right]  \left[
%TCIMACRO{\dprod _{i=2}^{l-1}}%
%BeginExpansion
{\displaystyle\prod_{i=2}^{l-1}}
%EndExpansion
\,\,%
%TCIMACRO{\dprod _{j=i+1}^{l}}%
%BeginExpansion
{\displaystyle\prod_{j=i+1}^{l}}
%EndExpansion
\operatorname{H}(D-\left\vert \mathbf{r}_{i}-\mathbf{\mathbf{r}}%
_{j}\right\vert )\right]
%TCIMACRO{\dprod _{i=2}^{l}}%
%BeginExpansion
{\displaystyle\prod_{i=2}^{l}}
%EndExpansion
\mathrm{d}^{3}\mathbf{r}_{i},\label{2.6}%
\end{equation}
\end{widetext}where
\begin{equation}
n(\mathbf{r},t)=\int f(\mathbf{v},\mathbf{r},t)\,\mathrm{d}^{3}\mathbf{v},
\label{2.7}%
\end{equation}
is the number density; $c_{2}$, $c_{3}$, ... $c_{L}$ are undetermined
coefficients; $\int^{l}$ denotes $l$ repeated integrals; and it is implied
that, if the lower limit of a product is larger than the upper one, the
product is equal to unity---e.g.,%
\[%
%TCIMACRO{\dprod _{i=2}^{1}}%
%BeginExpansion
{\displaystyle\prod_{i=2}^{1}}
%EndExpansion
\cdots=1.
\]
As shown in Ref. \cite{BenilovBenilov18}, series (\ref{2.6}) satisfies all the
fundamental requirements. The fluid of hard spheres correspond to $c_{l}=1$
for all $l$, and the individual terms in (\ref{2.6}) represent overlaps
between two, three, four, etc. spheres, to be excluded from the allowable
phase space. In the general case, the coefficients $c_{l}$ and the molecular
diameter $D$ should be treated as adjustable parameters, to be fixed when the
EV model is calibrated.

Let $\mathbf{F}$ in the kinetic equation (\ref{2.1}) be the van der Waals
force exerted by the molecules on one another, characterized by a pairwise
potential $\Phi(r)$ where $r$ is the distance between two molecules. The
collective force acting on a particle located at a position $\mathbf{r}$ can
then be written in the form%
\begin{equation}
\mathbf{F(r},t)=-\mathbf{\nabla}\int\Phi\left(  \left\vert \mathbf{r}%
-\mathbf{r}_{1}\right\vert \right)  \,n\mathbf{(r}_{1},t)\,\mathrm{d}%
^{3}\mathbf{r}_{1},\label{2.8}%
\end{equation}
where it is assumed that $\Phi(r)$ decays sufficiently fast as $r\rightarrow
\infty$. There are two global characteristics of $\Phi(r)$ that are
particularly important:%
\begin{equation}
a=-\frac{1}{2m^{2}}\int\Phi(r)\,\mathrm{d}^{3}\mathbf{r},\label{2.9}%
\end{equation}%
\begin{equation}
K=-\frac{1}{2m^{2}}\int\frac{r^{2}}{3}\Phi(r)\,\mathrm{d}^{3}\mathbf{r}%
,\label{2.10}%
\end{equation}
which will be referred to as the van der Waals and Korteweg constants,
respectively. The former is the same $a$ which appears in van der Waals's
classical equation of state, and the latter characterizes the fluid's surface
tension. Note also that%
\begin{equation}
R=\sqrt{\frac{K}{a}}\label{2.11}%
\end{equation}
can be viewed as a characteristic radius of the van der Waals force.

Since the potential of the `real' van der Waals force is difficult to
calculate, $\Phi(r)$ will be treated in this paper as an adjustable parameter.

\subsection{Energy and entropy\label{sec2.2}}

Assume for the sake of argument that $f(\mathbf{v},\mathbf{r},t)$ is spatially
periodic along all three dimensions. Then, it can be readily verified that the
EV kinetic model (\ref{2.1})--(\ref{2.8}) conserves the total number of
molecules and the total energy,%
\begin{equation}
N=\int n(\mathbf{r},t)\,\mathrm{d}^{3}\mathbf{r}, \label{2.12}%
\end{equation}%
\begin{multline}
E=%
%TCIMACRO{\diint }%
%BeginExpansion
{\displaystyle\iint}
%EndExpansion
\frac{m\left\vert \mathbf{v}\right\vert ^{2}}{2}f(\mathbf{v},\mathbf{r}%
,t)\,\mathrm{d}^{3}\mathbf{v}\,\mathrm{d}^{3}\mathbf{r}\\
+\frac{1}{2}%
%TCIMACRO{\diint }%
%BeginExpansion
{\displaystyle\iint}
%EndExpansion
n(\mathbf{r},t)\,n(\mathbf{r}_{1},t)\,\Phi\left(  \left\vert \mathbf{r}%
-\mathbf{r}_{1}\right\vert \right)  \,\mathrm{d}^{3}\mathbf{r}_{1}%
\mathrm{d}^{3}\mathbf{r}, \label{2.13}%
\end{multline}
where the integrals are implied to be over one cell of the periodic formation.
The total momentum is also conserved, but it is not needed in this paper.

As shown in Ref. \cite{GrmelaGarciacolin80}, the EV model satisfies an H
theorem---i.e., $\mathrm{d}S/\mathrm{d}t\geq0$, with the entropy integral
given by%
\begin{equation}
S=-%
%TCIMACRO{\diint }%
%BeginExpansion
{\displaystyle\iint}
%EndExpansion
f(\mathbf{r},\mathbf{v},t)\ln\frac{f(\mathbf{r},\mathbf{v},t)}{f_{0}%
}\mathrm{d}^{3}\mathbf{v}\,\mathrm{d}^{3}\mathbf{r}-k_{B}Q[n],\label{2.14}%
\end{equation}
where $k_{B}$ is the Boltzmann constant, $f_{0}$ is an arbitrary constant (can
be fixed as convenient), and the non-ideal part of the entropy, $Q$, is
related to the coefficient $\eta$ by\begin{widetext}%
\begin{equation}
\mathbf{\nabla}\frac{\delta Q[n]}{\delta n(\mathbf{r},t)}=\int\eta
(\mathbf{r},\mathbf{r}_{1},t)\,n(\mathbf{r}_{1},t)\,(\mathbf{r}_{1}%
-\mathbf{\mathbf{r}})\,\delta(\left\vert \mathbf{r}-\mathbf{\mathbf{r}}%
_{1}\right\vert -1)\,\mathrm{d}^{3}\mathbf{r}_{1}.\label{2.15}%
\end{equation}
If $\eta$ is given by expression (\ref{2.6}), Eq. (\ref{2.15}) yields \cite{BenilovBenilov18},%
\begin{multline}
Q[n]=\frac{1}{2}%
%TCIMACRO{\diint }%
%BeginExpansion
{\displaystyle\iint}
%EndExpansion
n(\mathbf{r},t)\,n(\mathbf{r}_{1},t)\,\operatorname{H}(D-\left\vert
\mathbf{r}-\mathbf{\mathbf{r}}_{1}\right\vert )\,\mathrm{d}^{3}\mathbf{r}%
\,\mathrm{d}^{3}\mathbf{r}_{1}\\
+\sum_{l=2}^{L}\frac{c_{l}}{l\left(  l+1\right)  }\int^{l}\int n(\mathbf{r}%
,t)\left[
%TCIMACRO{\dprod _{i=1}^{l}}%
%BeginExpansion
{\displaystyle\prod_{i=1}^{l}}
%EndExpansion
n(\mathbf{r}_{i},t)\operatorname{H}(D-\left\vert \mathbf{r}-\mathbf{\mathbf{r}%
}_{i}\right\vert )\right]  \left[
%TCIMACRO{\dprod _{i=1}^{l-1}}%
%BeginExpansion
{\displaystyle\prod_{i=1}^{l-1}}
%EndExpansion
\,\,%
%TCIMACRO{\dprod _{j=i+1}^{l}}%
%BeginExpansion
{\displaystyle\prod_{j=i+1}^{l}}
%EndExpansion
\operatorname{H}(D-\left\vert \mathbf{r}_{i}-\mathbf{\mathbf{r}}%
_{j}\right\vert )\right]  \mathrm{d}^{3}\mathbf{r\,}%
%TCIMACRO{\dprod _{i=1}^{l}}%
%BeginExpansion
{\displaystyle\prod_{i=1}^{l}}
%EndExpansion
\mathrm{d}^{3}\mathbf{r}_{i}.\label{2.16}%
\end{multline}
\end{widetext}Eq. (\ref{2.15}) also explains why Enskog's original collision
integral \cite{Enskog22} does not satisfy the H theorem: it assumes $\eta$ to
be a function of the density $n(\mathbf{r},t)$ evaluated at the impact point
of the colliding molecules---in which case Eq. (\ref{2.15}) admits solutions
for $Q$ only if $n$ satisfies an additional constraint\footnote{To obtain this
constraint, substitute $\eta=\eta(n((\mathbf{r}+\mathbf{r}_{1})/2,t))$ into
Eq. (\ref{2.15}) and take curl of both sides of the resulting equality.}.

\subsection{Equilibrium solutions of the EV equation\label{sec2.3}}

The present work is concerned with equilibrium states, in which case the
general EV model (\ref{2.1})--(\ref{2.8}) reduces to a much simpler equation.
This reduction exploits the fact that the scattering cross-section for hard
spheres, (\ref{2.4}), depends on the direction of $\mathbf{g}=\mathbf{v}%
_{1}-\mathbf{v}$, but not on $\left\vert \mathbf{g}\right\vert $; as a result,
the EV equation admits Maxwellian solutions with spatially uniform
temperature, but variable density,%
\begin{equation}
f=\frac{m^{3/2}}{\left(  2\pi k_{B}T\right)  ^{3/2}}n(\mathbf{r})\exp\left(
-\frac{mv^{2}}{2k_{B}T}\right)  , \label{2.17}%
\end{equation}
where $n(\mathbf{r})$ satisfies the following equation \cite{BenilovBenilov18}%
\begin{multline}
\ln n(\mathbf{r})+\frac{\delta Q[n]}{\delta n(\mathbf{r})}\\
+\frac{1}{k_{B}T}\int\Phi\left(  \left\vert \mathbf{r}-\mathbf{r}%
_{1}\right\vert \right)  \,n\mathbf{(r}_{1})\,\mathrm{d}^{3}\mathbf{r}%
_{1}=\operatorname{const}. \label{2.18}%
\end{multline}
Note that $\operatorname{const}$ does not depending on $\mathbf{r}$, but may
depend on $T$.

To understand the physical meaning of Eq. (\ref{2.18}), note that it follows
from the maximum entropy principle.

Indeed, substituting the Maxwellian ansatz (\ref{2.17}) into expressions
(\ref{2.12})--(\ref{2.14}), one obtains%
\begin{equation}
N=\int n(\mathbf{r})\,\mathrm{d}^{3}\mathbf{r}, \label{2.19}%
\end{equation}%
\begin{multline}
E=\frac{3}{2}k_{B}T\int n(\mathbf{r})\,\mathrm{d}^{3}\mathbf{r}\\
+\frac{1}{2}%
%TCIMACRO{\diint }%
%BeginExpansion
{\displaystyle\iint}
%EndExpansion
\Phi\left(  \left\vert \mathbf{r}-\mathbf{r}_{1}\right\vert \right)
\,n(\mathbf{r})\,n(\mathbf{r}_{1})\,\mathrm{d}^{3}\mathbf{r}_{1}\mathrm{d}%
^{3}\mathbf{r}, \label{2.20}%
\end{multline}%
\begin{multline}
S=\frac{3}{2}k_{B}\int n(\mathbf{r})\left[  \ln\frac{2\pi k_{B}T\,f_{0}^{2/3}%
}{m\,n^{2/3}(\mathbf{r})}+1\right]  \mathrm{d}^{3}\mathbf{r}\\
-k_{B}Q[n]. \label{2.21}%
\end{multline}
It can now be verified that Eq. (\ref{2.18}) is equivalent to the requirement
that the entropy be at an extremum, subject to constraints of constant energy
and number of molecules---i.e.,%
\begin{equation}
\delta\left(  S+\lambda E+\mu N\right)  =0, \label{2.22}%
\end{equation}
where $\lambda$ and $\mu$ are the Lagrange multipliers. Substituting
expressions (\ref{2.19})--(\ref{2.21}) into Eq. (\ref{2.22}) and equating to
zero the coefficients of $\delta T$ and $\delta n(\mathbf{r})$, one can first
show that $\lambda=-1/T$, and then recover Eq. (\ref{2.18}) with
$\operatorname{const}=\mu/k_{B}T-1$.

\subsection{Thermodynamics of EV fluids\label{sec2.4}}

The Enskog--Vlasov kinetic model implies certain thermodynamic properties of
the fluid under consideration. These properties can be conveniently deduced
from the EV\ energy and entropy integrals using the formalism proposed in Ref.
\cite{GiovangigliMatuszewski13}.

\subsubsection{Thermodynamic definitions\label{sec2.4.1}}

According to Ref. \cite{GiovangigliMatuszewski13}, a fluid is fully
characterized by the dependence of its internal energy $e$ and entropy $s$
(both per molecule) on the number density $n$ and temperature $T$. Note that
$e(n,T)$ and $s(n,T)$ are supposed to satisfy the Gibbs relation, whose
differential form amounts to \cite{Benilov23a}%
\begin{equation}
\frac{\partial e}{\partial T}=T\frac{\partial s}{\partial T}.\label{2.23}%
\end{equation}
If $e(n,T)$ and $s(n,T)$ are known for the fluid under consideration, its
pressure $p(n,T)$ and chemical potential $G(n,T)$ are
\cite{GiovangigliMatuszewski13}%
\begin{equation}
p=n^{2}\left(  \frac{\partial e}{\partial n}-T\frac{\partial s}{\partial
n}\right)  ,\label{2.24}%
\end{equation}%
\begin{equation}
G=\frac{\partial(ne)}{\partial n}-T\frac{\partial(ns)}{\partial n}%
.\label{2.25}%
\end{equation}
Using $p(n,T)$ and $G(n,T)$, one can calculate the liquid--vapor saturation
curve through the so-called Maxwell construction---i.e., the conditions%
\begin{equation}
p(n^{(v)},T)=p(n^{(l)},T),\label{2.26}%
\end{equation}%
\begin{equation}
G(n^{(v)},T)=G(n^{(l)},T),\label{2.27}%
\end{equation}
where $n^{(v)}$ and $n^{(l)}$ are the vapor and liquid densities, respectively.

\subsubsection{Pressure and chemical potential of EV fluids\label{sec2.4.2}}

To determine the per-molecule internal energy $e$ and entropy $s$ of an EV
fluid, one needs to apply the thermodynamic approximation to the total EV
energy, (\ref{2.20}), and the total EV entropy, (\ref{2.21}). Assuming, thus,
that the spatial scale of $n(\mathbf{r})$ exceeds both the molecular diameter
$D$ and the radius $R$ of the van der Waals force, one can represent $E$ and
$S$ in the form%
\[
E=\int n\,e\,\mathrm{d}^{3}\mathbf{r},\qquad S=\int n\,s\,\mathrm{d}%
^{3}\mathbf{r},
\]
with%
\begin{equation}
e=C_{V}T-am^{2}n,\label{2.28}%
\end{equation}%
\begin{equation}
s=C_{V}\ln T-k_{B}\ln(D^{3}n)-k_{B}\theta(D^{3}n)+\cdots,\label{2.29}%
\end{equation}
where $C_{V}=3k_{B}/2$ is the heat capacity at fixed volume, $a$ is the van
der Waals constant (\ref{2.9}), and $\cdots$ hides the terms that include
neither $n$ nor $T$ (and thus do not contribute to Eqs. (\ref{2.23}%
)--(\ref{2.25}) and, in any case, can be eliminated by an appropriate choice
of $f_{0}$). The function $\theta(\xi)$ is given by%
\begin{equation}
\theta(\xi)=\frac{2\pi}{3}n_{nd}+\sum_{l=2}^{L}\frac{c_{l}A_{l}}{l\left(
l+1\right)  }\xi^{l},\label{2.30}%
\end{equation}
where the numeric constants $A_{l}$ ($l\geq2$) are%
\begin{multline}
A_{l}=\int^{l}\left[
%TCIMACRO{\dprod _{i=1}^{l}}%
%BeginExpansion
{\displaystyle\prod_{i=1}^{l}}
%EndExpansion
\operatorname{H}(1-\left\vert \mathbf{\mathbf{r}}_{i}\right\vert )\right]  \\
\times\left[
%TCIMACRO{\dprod _{i=1}^{l-1}}%
%BeginExpansion
{\displaystyle\prod_{i=1}^{l-1}}
%EndExpansion
\,\,%
%TCIMACRO{\dprod _{j=i+1}^{l}}%
%BeginExpansion
{\displaystyle\prod_{j=i+1}^{l}}
%EndExpansion
\operatorname{H}(1-\left\vert \mathbf{r}_{i}-\mathbf{\mathbf{r}}%
_{j}\right\vert )\right]
%TCIMACRO{\dprod _{i=1}^{l}}%
%BeginExpansion
{\displaystyle\prod_{i=1}^{l}}
%EndExpansion
\mathrm{d}^{3}\mathbf{r}_{i}.\label{2.31}%
\end{multline}
$A_{2}$ can be calculated analytically, and $A_{3}$, $A_{4}$, $A_{5}$... can
be computed numerically, using the Monte Carlo method:%
\[
A_{2}=\frac{5\pi^{2}}{6},
\]%
\[
A_{3}\approx11.6396,\qquad A_{4}\approx13.6845,\qquad A_{5}\approx14.2689...
\]
Substituting (\ref{2.28})--(\ref{2.29}) into (\ref{2.24})--(\ref{2.25}), one
obtains the following expressions for the pressure and chemical potential:%
\begin{equation}
p=k_{B}Tn\left[  1+D^{3}n\,\theta^{\prime}(D^{3}n)\right]  -am^{2}%
n^{2},\label{2.32}%
\end{equation}%
\begin{multline}
G=k_{B}T\left[  \ln D^{3}n+1+\theta(D^{3}n)+D^{3}n\,\theta^{\prime}%
(D^{3}n)\right]  \\
-2am^{2}n,\label{2.33}%
\end{multline}
where $\theta^{\prime}(\xi)=\mathrm{d}\theta(\xi)/\mathrm{d}\xi$ and
$\theta(\xi)$ is given by (\ref{2.30})--(\ref{2.31}).

\subsection{Calibration of the Enskog-Vlasov model\label{sec2.5}}

The EV model includes the following adjustable parameters: the molecular
diameter $D$, the coefficients $c_{l}$ ($l\geq2$), and the van der Waals
potential $\Phi(r)$. They are to be determined by fitting the properties of
the `theoretical' EV fluid to those of the `real' fluid under consideration.

Three comments are in order.

\begin{itemize}
\item Series (\ref{2.6}), (\ref{2.16}), and (\ref{2.30}) are truncated at
$L=5$. Such an approximation is sufficient for inert fluids (see Refs.
\cite{BenilovBenilov18,BenilovBenilov19a,BenilovBenilov19b}) and it is
comparably accurate at least for some diatomic fluids (Appendix \ref{appA} of
the present paper). It should also be noted that the numerical solution of the
EV equation involves the evaluation of $(2L-1)$-order integrals, making it
essential to keep $L$ at the smallest sufficient value.

\item As shown in Ref. \cite{BenilovBenilov18}, the fitted values of $D$ for
inert fluids are remarkably close to%
\begin{equation}
D=\left(  \frac{m}{\rho_{tp}^{(l)}}\right)  ^{1/3},\label{2.34}%
\end{equation}
where $\rho_{tp}^{(l)}$ is the mass density of the corresponding liquid at the
triple point. It appears that expression (\ref{2.34}) is the natural choice
for the molecular diameter, and it is used in the present paper for all fluids
without adjustment.

\item As argued in Ref. \cite{BenilovBenilov19a}, the solutions of the EV
equation are not sensitive to the specific shape of $\Phi(r)$, but rather
depend on the van der Waals and Korteweg constants, $a$ and $K$, given by
(\ref{2.9})--(\ref{2.10}), respectively.\newline\hspace*{0.5cm}Several
examples of $\Phi(r)$ characterized by the same values of $a$ and $K$ have
been tested, and the conclusion of Ref. \cite{BenilovBenilov19a} that the
actual shape of $\Phi(r)$ is unimportant has been confirmed. The results
presented in this paper were computed for the following particular case:%
\begin{equation}
\Phi=-\frac{2m^{2}a}{\left(  2\pi\right)  ^{3/2}R^{3}}\exp\left(  -\frac
{r^{2}}{2R^{2}}\right)  , \label{2.35}%
\end{equation}
where the characteristic radius $R$ of the van der Waals force is related to
$a$ and $K$ through (\ref{2.11}).
\end{itemize}

Further details of the calibration of the EV model are explained in Appendix
\ref{appA}. The fitted parameter values for the fluids considered are
presented in Table \ref{tab1}. Note that the constants $a$ and $K$ of water
exceed those of the other fluids, implying a stronger van der Waals force
(likely due to hydrogen bonding).

\begin{table*}
\begin{ruledtabular}\begin{tabular}{lcccccccc}
\rule{0pt}{4.5mm} \vspace{1.5mm} & $c_{2}$ & $c_{3}$ & $c_{4}$ & $c_{5}$ & $a~(\mathrm{m}^{5}/\mathrm{s}^{2}\,\mathrm{kg})$
& $K~(10^{-18}\mathrm{m}^{7} /\mathrm{s}^{2}\,\mathrm{kg})$ & $\frac{1}{2}D\,(\mathrm{\mathring{A}})$ & $R\,(\mathrm{\mathring{A}})$\\
\hline
\rule{0pt}{4mm} $\mathrm{Ar}$\vspace{1mm} & $-1.717894$ & $9.350866$ & $-12.498816$ & $6.114199$ & $\;\;101.99$ & $\;\;3.39$ & $1.80$ & $1.82$\\
\rule{0pt}{4mm} $\mathrm{N}_{2}$\vspace{1mm} & $\;\;\;0.077744$ & $6.321150$ & $-10.323002$ & $5.872395$ & $\;\;222.24$ & $\;\;8.71$ & $1.89$ & $1.98$\\
\rule{0pt}{4mm} $\mathrm{O}_{2}$\vspace{1mm} & $\;\;\;1.016127$ & $6.227040$ & $-12.847245$ & $8.710581$ & $\;\;172.73$ & $\;\;6.53$ & $1.72$ & $1.94$\\
\rule{0pt}{4mm} $\mathrm{F}_{2}$\vspace{1mm} & $\;\;\;1.469203$ & $5.467670$ & $-12.172963$ & $8.354169$ & $\;\;106.38$ & $\;\;3.72$ & $1.67$ & $1.87$\\
\rule{0pt}{4mm} $\mathrm{CO}$\vspace{1mm} & $\;\;\;0.422063$ & $5.461269$ & $\;\;-9.273882$ & $5.388458$ & $\;\;242.17$ & $\;\;9.71$ & $1.90$ & $2.00$\\
\rule{0pt}{4mm} $\mathrm{H}_{2}\mathrm{O}$\vspace{1mm} & $\;\;\;4.648588$ & $1.643674$ & $-10.109812$ & $7.973337$ & $2112.10$ & $69.60$ & $1.55$ & $1.82$
\end{tabular}\end{ruledtabular}
\caption{The fitted values of the coefficients $c_{l}$, the van der Waals parameter $a$, the Korteweg parameter $K$, the molecular radius $D/2$, and the radius $W$ of the van der Waals force.}
\label{tab1}
\end{table*}

It is also interesting to compare the radius $R$ of the van der Waals force,
(\ref{2.11}), with the molecular radius, $D/2$ (see the last two columns of
Table \ref{tab1}). Evidently, the former parameter consistently exceeds the
latter, but not by much. Thus, the van der Waals force can hardly be regarded
as long-range (which is how it is usually described in textbooks).

\section{Liquid--vapor interfaces\label{sec3}}

\subsection{Formulation\label{sec3.1}}

Let a flat horizontal interface be characterized by a density profile $n(z)$,
where $z$ is the vertical coordinate. Rewriting Eq. (\ref{2.18}) and
expression (\ref{2.16}) in cylindrical coordinates $\left(  x,y,z\right)
\rightarrow\left(  r_{\bot},\phi,z\right)  $, $\left(  x_{1},y_{1}%
,z_{1}\right)  \rightarrow\left(  r_{1\bot},\phi_{1},z_{1}\right)  $, etc.,
one obtains\begin{widetext}%
\begin{multline}
\ln n(z)+\pi\int_{-D}^{D}n(z_{1}-z)\,\left(  D^{2}-z_{1}^{2}\right)
\,\mathrm{d}z_{1}+\sum_{j=2}^{L}\frac{c_{l}}{l}\left(  \int_{-\infty}^{\infty
}\right)  ^{l}\left[
%TCIMACRO{\dprod _{j=1}^{l}}%
%BeginExpansion
{\displaystyle\prod_{j=1}^{l}}
%EndExpansion
n(z_{j}-z)\right]  F_{l}(z_{1},z_{2}...z_{l})%
%TCIMACRO{\dprod _{j=1}^{l}}%
%BeginExpansion
{\displaystyle\prod_{j=1}^{l}}
%EndExpansion
\mathrm{d}z_{j}\\
+\frac{1}{k_{B}T}\int_{-\infty}^{\infty}n(z_{1}-z)\,\Psi(z_{1})\,\mathrm{d}%
z_{1}=\operatorname{const},\label{3.1}%
\end{multline}
where the kernel functions $F_{l}$ ($l\geq2$) are given by%
\begin{multline}
F_{l}=2\pi\int_{0}^{\infty}\left(  \int_{0}^{2\pi}\int_{0}^{\infty}\right)
^{l-1}\left[
%TCIMACRO{\dprod _{j=1}^{l}}%
%BeginExpansion
{\displaystyle\prod_{j=1}^{l}}
%EndExpansion
\operatorname{H}\left(  D^{2}-r_{j\bot}^{2}-z_{j}^{2}\right)  \right]  \left[
%
%TCIMACRO{\dprod _{j=1}^{l-1}}%
%BeginExpansion
{\displaystyle\prod_{j=1}^{l-1}}
%EndExpansion
\operatorname{H}\left(  D^{2}-r_{j\bot}^{2}+2r_{j\bot}r_{l\bot}\cos\phi
_{j}-r_{l\bot}^{2}-\left(  z_{j}-z_{l}\right)  ^{2}\right)  \right]  \\
\times\left[
%TCIMACRO{\dprod _{j=1}^{l-2}}%
%BeginExpansion
{\displaystyle\prod_{j=1}^{l-2}}
%EndExpansion%
%TCIMACRO{\dprod _{i=j+1}^{l-1}}%
%BeginExpansion
{\displaystyle\prod_{i=j+1}^{l-1}}
%EndExpansion
\operatorname{H}\left(  D^{2}-r_{j\bot}^{2}+2r_{j\bot}r_{i\bot}\cos\left(
\phi_{j}-\phi_{i}\right)  -r_{i\bot}^{2}-\left(  z_{j}-z_{i}\right)
^{2}\right)  \right]  \left[
%TCIMACRO{\dprod _{j=1}^{l-1}}%
%BeginExpansion
{\displaystyle\prod_{j=1}^{l-1}}
%EndExpansion
r_{j\bot}\mathrm{d}r_{j\bot}\mathrm{d}\phi_{j}\right]  r_{l\bot}%
\mathrm{d}r_{l\bot},\label{3.2}%
\end{multline}
\end{widetext}and%
\begin{equation}
\Psi(z)=2\pi\int_{0}^{\infty}\Phi\left(  \sqrt{r_{\bot}^{2}+z^{2}}\right)
\,r_{\bot}\mathrm{d}r_{\bot},\label{3.3}%
\end{equation}
is the one-dimensional reduction of the van der Waals potential, with
$\Phi(r)$ given by (\ref{2.35}).

To single out the solution of Eq. (\ref{3.1}) describing an interface, require%
\begin{equation}
n\rightarrow n^{(l)}\qquad\text{as}\qquad z\rightarrow-\infty, \label{3.4}%
\end{equation}%
\begin{equation}
n\rightarrow n^{(v)}\qquad\text{as}\qquad z\rightarrow+\infty, \label{3.5}%
\end{equation}
where $n^{(l)}$ and $n^{(v)}$ are the liquid and vapor densities,
respectively. Note that they must satisfy the Maxwell construction,
(\ref{2.26})--(\ref{2.27}), otherwise boundary-value problem (\ref{3.1}%
)--(\ref{3.5}) would not have a solution \cite{BenilovBenilov18}.

Boundary conditions (\ref{3.4})--(\ref{3.5}) can also be used to find the
constant on the right-hand side of Eq. (\ref{3.1}) without solving the whole
problem: taking in (\ref{3.1}) the limit $z\rightarrow-\infty$, one can show
that condition (\ref{3.4}) implies that $\operatorname{const}=G(n^{(l)}%
,T)/k_{B}T-1$, where $G(n,T)$ is the chemical potential given by (\ref{2.33}).
The limit $z\rightarrow+\infty$ and condition (\ref{3.5}), in turn, yield
$\operatorname{const}=G(n^{(v)},T)/k_{B}T-1$, which is the same value due to
requirement (\ref{2.27}) of the Maxwell construction.

Since Eq. (\ref{3.1}) and boundary conditions (\ref{3.4})--(\ref{3.5}) are
translationally invariant, the solution $n(z)$ can be shifted by an arbitrary
distance along the $z$ axis and would still remain a solution. To make it
unique, one can fix $n$ at a single point, for example,%
\begin{equation}
n=\frac{1}{2}\left(  n^{(l)}+n^{(v)}\right)  \qquad\text{at}\qquad
z=0.\label{3.6}%
\end{equation}
Once the solution $n(z)$ of boundary-value problem (\ref{3.1})--(\ref{3.6}) is
obtained, the surface tension can be calculated using the formula
\cite{BenilovBenilov18,Benilov24a}%
\[
\gamma=\int_{-\infty}^{\infty}\int_{-\infty}^{\infty}n(z)\,n(z_{1}%
)\,\Gamma(z-z_{1})\,\mathrm{d}z_{1}\mathrm{d}z,
\]
where%
\[
\Gamma(z)=\frac{1}{2}\Psi(z)-\pi z^{2}\Phi(z).
\]
Eq. (\ref{3.1}) is a nonlinear integral equation with no small parameters, so
neither exact nor asymptotic solutions are available. The only option is
numerical computation, but even this is not straightforward: expressions
(\ref{3.2}) involve $2l-1$ integrations, making the kernels $F_{l}$
increasingly difficult to compute. There is a mitigating circumstance,
however: if (\ref{3.2}) is rewritten in terms of%
\[
z_{1nondim}=\frac{z_{1}}{D},\qquad z_{2nondim}=\frac{z_{2}}{D},\qquad\cdots,
\]
it become parameterless---hence, once computed, it can be used for \emph{all}
fluids under consideration. It is also helpful that all $F_{l}(z_{1}...z_{l})$
vanish if one of their arguments lies outside the interval $\left(
-D,D\right)  $.

The numerical algorithm used for solving boundary-value problem (\ref{3.1}%
)--(\ref{3.6}) is described in Appendix \ref{appB}.

\subsection{Results and discussion\label{sec3.2}}

\begin{figure}
\includegraphics[width=\columnwidth]{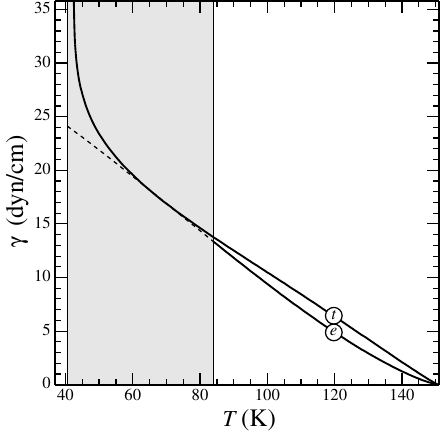}
\caption{The surface tension $\gamma$ vs. the temperature $T$, for argon. The supercooling region, between the spinodal and triple points, is shaded. Curve (t) is computed theoretically, using the EV model (\ref{3.1})--(\ref{3.6}); curve (e) corresponds to the empirical formula of Ref. \cite{Somayajulu88}. The latter strictly applies to the range between the triple and critical points, so its extension into the supercooling region is shown by a dotted line.}
\label{fig1}
\end{figure}

Fig. \ref{fig1} was computed for argon. One can see that, when the temperature
approaches a certain threshold (which will be referred to as spinodal and
denoted by $T_{s}$), the surface tension $\gamma$ appears to tend to infinity.

Three comments are in order:

\begin{itemize}
\item No solutions describing liquid--vapor interface have been found for
subspinodal temperatures, $T\leq T_{s}$.

\item The reason of the singularity of $\gamma$ is clarified by Fig.
\ref{fig2}, illustrating how the interfacial profile changes when the
temperature approaches the spinodal value. As $T\rightarrow T_{s}$, a region
forms in the liquid near the interface that becomes `filled' with oscillations
of increasing amplitude. Since surface tension measures interfacial energy,
its divergence is unsurprising.

\begin{figure*}
\includegraphics[width=\textwidth]{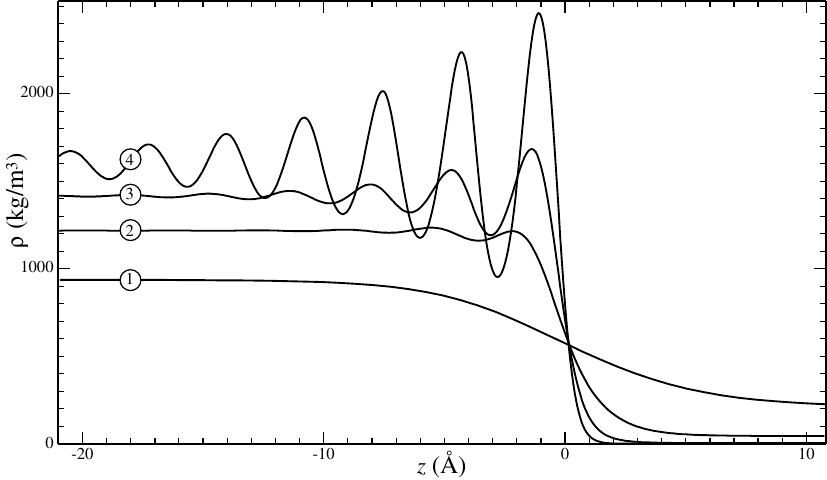}
\caption{Spatial structure of liquid--vapor interfaces in argon: $\rho=mn$ is the mass density, $z$ is the coordinate. The temperatures are: (1) $T=145\,\mathrm{K}$, (2) $T=117\,\mathrm{K}$, (3) $T=84\,\mathrm{K}$, (4) $T=55\,\mathrm{K}$. For comparison, $T_{cr}\approx151\,\mathrm{K}$, $T_{tp}\approx84\,\mathrm{K}$, and $T_{s}\approx40.6\,\mathrm{K}$.}
\label{fig2}
\end{figure*}

\item As shown in the next section, $T_{s}$ is the temperature at which liquid
becomes unstable and crystallizes. The oscillation emerging in the interfacial
profile as $T\rightarrow T_{s}$ is a precursor to the disturbance that becomes
unstable at $T=T_{s}$. At subspinodal temperatures, this disturbance
\emph{decays} while propagating into the liquid, but then the \emph{decay rate
vanishes}, and the disturbance extends to infinity.
\end{itemize}

The pattern illustrated in Figs. \ref{fig1}--\ref{fig2} for argon was
consistently observed for all examined fluids ($\mathrm{N}_{2}$,
$\mathrm{O}_{2}$, $\mathrm{F}_{2}$, $\mathrm{CO}$, and $\mathrm{H}%
_{2}\mathrm{O}$): for each of them, both the surface tension and the width of
the oscillatory region diverge as $T\rightarrow T_{s}$ . This is not
surprising, as the spinodal singularity of surface tension has a clear
physical interpretation and should therefore exist for all fluids,
\emph{regardless of the model used to describe them}. After all, any adequate
model should describe crystallization---hence, be unstable with respect to a
periodic wave emerging when the spinodal point is reached.

Note also the discrepancy between the theoretical and empirical curves in the
near-critical region of Fig. \ref{fig1}: the surface tension is so small there
that it is comparable to the overall inaccuracy of the EV model. This region,
however, is not of interest for the present paper, which focuses on supercooling.

\section{Spinodal instability\label{sec4}}

As shown in Ref. \cite{BenilovBenilov19a}, the EV model predicts two types of
cooling-induced spinodal instability: a transition from one fluid state to
another (vapor to liquid) and a transition from fluid to solid (liquid to
crystal). The former occurs at an infinitely large spatial scale, whereas the
latter occurs at a finite scale equal to the period of the emerging
crystalline structure. These results were illustrated for inert fluids, using
a partially calibrated model, in which $c_{l}$ and $a$ were fixed, but $K$ was
not. Consequently, the spinodal temperature at which crystallization occurs
was not calculated.

This task has now been completed, and the results are presented below.

\subsection{Formulation\label{sec4.1}}

As in Ref. \cite{BenilovBenilov19a}, the fluid stability in this work is
analyzed using \textquotedblleft frozen waves\textquotedblright, i.e. harmonic
disturbances with zero growth rate. These can be viewed as precursors of
unstable disturbances: if a frozen wave with a wavenumber $k$ exists for a
given state, then a small increase or decrease in $k$ renders the state
unstable---and the temperature at which the first frozen wave arises marks the
onset of instability.

At the same time, frozen waves are one-dimensional and stationary---thus,
described by the steady equation (\ref{3.1}), which is incomparably simpler
than the full EV model (\ref{2.1})--(\ref{2.8}). The latter is what one would
have to use if studying growing disturbances.

Consider a spatially uniform state of density $\bar{n}$ and let%
\begin{equation}
n=\bar{n}+\tilde{n}(z), \label{4.1}%
\end{equation}
where $\tilde{n}(z)$ is a small perturbation. Substituting (\ref{4.1}) into
Eq. (\ref{3.1}), linearizing it, letting the perturbation be harmonic,%
\[
\tilde{n}(z)=\operatorname{e}^{\mathrm{i}kz},
\]
and omitting the overbars ($\bar{n}\rightarrow n$), one obtains, after
straightforward algebra,\begin{widetext}%
\begin{equation}
k_{B}T=-\frac{n\Psi(k)}{1+\dfrac{4\pi}{k^{3}}\left(  \sin Dk-Dk\cos Dk\right)
n+%
%TCIMACRO{\dsum \limits_{l=2}^{L}}%
%BeginExpansion
{\displaystyle\sum\limits_{l=2}^{L}}
%EndExpansion
c_{l}n^{l}\hat{F}_{l}(k)},\label{4.2}%
\end{equation}
where%
\begin{equation}
\hat{\Psi}(k)=\int_{-\infty}^{\infty}\Psi(z_{1})\operatorname{e}%
^{\mathrm{i}kz_{1}}\mathrm{d}z_{1},\label{4.3}%
\end{equation}%
\begin{equation}
\hat{F}_{l}(k)=\int_{-\infty}^{\infty}\left[  \left(  \int_{-\infty}^{\infty
}\right)  ^{l-1}F_{l}(z_{1},z_{2}...z_{l})%
%TCIMACRO{\dprod _{j=1}^{l-1}}%
%BeginExpansion
{\displaystyle\prod_{j=1}^{l-1}}
%EndExpansion
\mathrm{d}z_{j}\right]  \operatorname{e}^{\mathrm{i}kz_{l}}\mathrm{d}%
z_{l}.\label{4.4}%
\end{equation}
\end{widetext}The function $\hat{\Psi}(k)$ is the Fourier transform of
$\Psi(z)$ given by (\ref{3.3}), and it can be readily calculated analytically.
The function $\hat{F}_{2}(k)$ can also be calculated, but $\hat{F}_{l}(k)$
with $l\geq3$ have to be computed numerically. Note that, even though
functions $\hat{F}_{l}(k)$ involve a parameter, the molecular diameter $D$, it
can be scaled out by representing $\hat{F}_{l}$ as a function of $Dk$ instead
of $k$, so one can compute them once and for all.

Eq. (\ref{4.2}) is a stability criterion: if, for a state $\left(  n,T\right)
$, (\ref{4.2}) admits a solution for some $k$, this state is unstable.

\subsection{Results and discussion\label{sec4.2}}

Eq. (\ref{4.2}) describes a family of curves $T$ vs. $n$, whose members are
characterized by different values of $k$ and are therefore denoted by
$T_{k}(n)$. The part of the $\left(  n,T\right)  $ plane swept by these curves
is the instability region, and its boundary is the spinodal curve.

\begin{figure}
\includegraphics[width=\columnwidth]{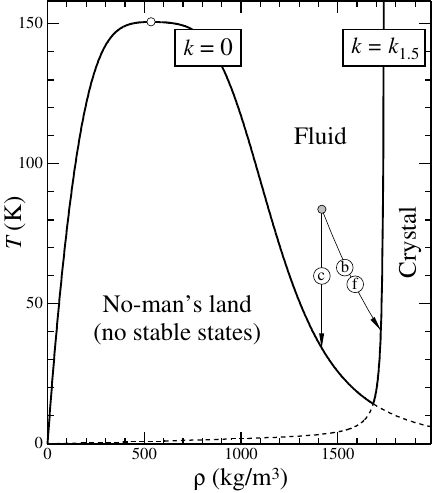}
\caption{Stability portrait of argon in the density-temperature plane. The curves labeled $k=0$ and $k=k_{1.5}$ correspond to spinodal instabilities with respect to infinitely-long and finite-length waves, respectively. Curves (c), (b), and (f)---the last two virtually coincident---show the isoChoric, isoBaric, and interFacial trajectories, respectively. The small empty circle denotes the critical point, the shaded circle denotes the triple point.}
\label{fig3}
\end{figure}

The typical behavior of the curves $T_{k}(n)$ is explained below and
illustrated for argon in Fig \ref{fig3}. Note that this and further diagrams
in this paper are drawn in the \emph{density}-temperature plane, not the
\emph{pressure}-temperature plane. The former choice conveniently allows one
to distinguish between states with the same pressure but different densities.

The behavior of the curve $T_{k}(n)$ when $k$ changes from $0$ to $\infty$,
and its implications for the liquid stability, are illustrated in Fig.
\ref{fig3} and explained below.

\begin{itemize}
\item The curve $T_{0}(n)$ is labeled in a self-explanatory manner. It can be
shown that its maximum is located at the critical point.

\item With increasing $k$, the curve $T_{k}(n)$ recedes, so that $T_{0}(n)$
separates the stability region above it from the instability region below it.
One can show that the latter corresponds to%
\[
\frac{\partial p(n,T)}{\partial n}<0.
\]
This condition predicts instability when a local density increase creates a
local pressure dip, thereby drawing in additional fluid.

\item If $k$ increases beyond a certain threshold (say, $k_{1}$), the
denominator of expression (\ref{4.2}) starts to vanish at a certain density
(say, $n_{\ast}$), where the corresponding curve $T(n)$ has a vertical
asymptote. At $k=k_{1}$, this asymptote enters through the infinity of the $n$
axis, and it retreats back to infinity when $k$ reaches a further threshold
(say, $k_{2}$).\newline\hspace*{0.5cm}Therefore, there must exist $k_{1.5}$
between $k_{1}$ and $k_{2}$, where $n_{\ast}$ reaches a minimum (say,
$n_{\ast1.5}$). The region of Fig. \ref{fig3} located to the right of the
curve $T_{k_{1.5}}(n)$ is unstable.

\item There exist further intervals $\left(  k_{3},k_{4}\right)  $, $\left(
k_{5},k_{6}\right)  $, etc., where the denominator of expression (\ref{4.2})
vanishes at a certain $n_{\ast}$. The corresponding minimum values of
$n_{\ast}$, however, are larger than $n_{\ast1.5}$, so no new regions of
instability arise.
\end{itemize}

Figure \ref{fig3} also shows examples of `trajectories' along which a liquid
could be supercooled: isobaric (at constant pressure), isochoric (at constant
volume), and what may be dubbed interfacial (i.e., in a closed container where
liquid and vapor are separated by a flat interface). For ease of comparison,
all three trajectories originate from the triple point.

Evidently, the isobaric and interfacial trajectories are virtually
indistinguishable. Observe also that

\begin{enumerate}
\item[(i)] the isobaric and interfacial trajectories are destabilized by a
finite-length wave, $k=k_{1.5}$;

\item[(ii)] whereas the isochoric trajectory is destabilized by an
infinitely-long wave, $k=0$.
\end{enumerate}

Once a trajectory crosses the spinodal curve, the liquid becomes unstable and
turns into crystal, but a difference should be expected between cases (i) and (ii).

In case (i), one may conjecture that the density of the resulting crystalline
argon is comparable to that of the pre-transition liquid, $\rho^{(l)}%
\approx1721\,\mathrm{kg}/\mathrm{m}^{3}$, and that the interatomic distance is
comparable to $2\pi/k\approx3.18\,\mathrm{\mathring{A}}$. For comparison, the
triple-point density of crystalline argon is $1616\,\mathrm{kg}/\mathrm{m}%
^{3}$, and the corresponding nearest-neighbor distance is
$3.75\,\mathrm{\mathring{A}}$.

In case (ii), the characteristics of the resulting crystal are likely
determined in the course of solidification. In particular, the crystal's
density should exceed noticeably that of the pre-transition liquid, since the
latter is well below the triple-point density of crystalline argon.

To emphasize the difference, the right-hand region of instability in Fig.
\ref{fig3} is labeled \textquotedblleft Crystal\textquotedblright\ (as it is
likely populated by crystalline states), and the lower region
\textquotedblleft No-man's land\textquotedblright\ (no stable states of any phase).

Observe also that Fig. \ref{fig3} does not include a separate region for
glassy states, consistent with the fact that polyamorphism has not been
observed for argon. For water, however, two glassy states are known to exist,
low- and high-density, thermodynamically distinct from each other and from the
liquid. Capturing such behavior requires an EV model that accounts for
molecular asymmetry and rotation, whereas the present version yields only
qualitative predictions.

\subsection{Comparison between different fluids\label{sec4.3}}

To facilitate comparison between $\mathrm{Ar}$, $\mathrm{N}_{2}$,
$\mathrm{O}_{2}$, $\mathrm{F}_{2}$, $\mathrm{CO}$, and $\mathrm{H}%
_{2}\mathrm{O}$, their stability portraits are shown in Fig. \ref{fig4} in
nondimensional variables. The following features can be observed:

\begin{figure*}
\includegraphics[width=\textwidth]{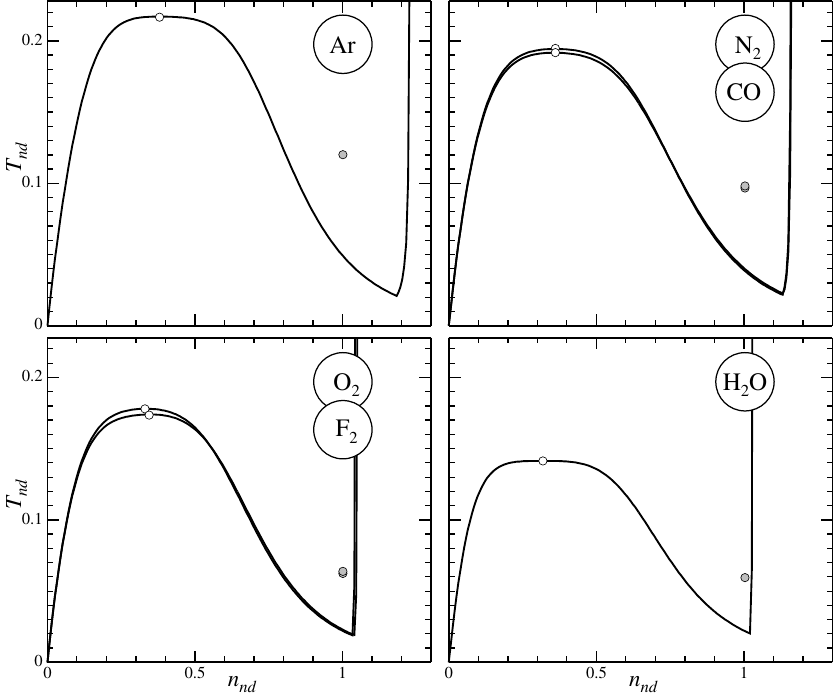}
\caption{The stability portraits of argon, nitrogen \& carbon monoxide, oxygen \& fluorine, and water, in terms of the nondimensional variables, $n_{nd}=D^{3}n$ and $T_{nd}=k_{B}T/am^{2}D^{3}$. The small empty circles show the fluids' critical points; the curves passing through them correspond to marginal stability with respect to infinitely long waves. The near-vertical curves correspond to instability at finite wavelengths. The shaded circles show the fluids' triple points.}
\label{fig4}
\end{figure*}

\begin{itemize}
\item The variation in the distance between the fluid's triple point and the
liquid--crystal separation curve reflects differences in the compressibility
of the liquid phase. Liquid argon is the most compressible, whereas water is
nearly incompressible, in agreement with intuition.

\item The evident similarity of the stability portraits of (quadrupolar)
nitrogen and (dipolar) carbon monoxide suggests that molecular polarity is not
a determining factor for the fluid's stability.

\item The similarity of $\mathrm{N}_{2}$ and $\mathrm{CO}$ cannot be
attributed solely to the near equality of their molecular weights, since the
portraits of $\mathrm{F}_{2}$ and $\mathrm{O}_{2}$ are also similar but their
molecular weights are different.

\item The weight difference between $\mathrm{F}_{2}$ and $\mathrm{O}_{2}$ is
the same as that between $\mathrm{O}_{2}$ and $\mathrm{N}_{2}$, yet the
portraits of the former pair are similar, whereas those of the latter differ
noticeably (compare the portrait of $\mathrm{O}_{2}$ in the bottom-left panel
of Fig. \ref{fig4} with that of $\mathrm{N}_{2}$ in the top-right panel).
\end{itemize}

Thus, there seems to be no consistent correlation between a fluid's stability
properties and its molecular weight or polarity. The only discernible
correlation which was found is that of $\Delta T=T_{tp}-T_{s}$ (the
temperature drop by which a fluid can be supercooled) for isobaric or
interfacial supercooling, on%
\begin{equation}
\kappa=\frac{R^{2}}{\left(  D/2\right)  ^{2}}, \label{4.5}%
\end{equation}
where $D/2$ is the molecular radius, and $R$ is the radius of the van der
Waals force given by (\ref{2.11}).

\begin{figure}
\includegraphics[width=\columnwidth]{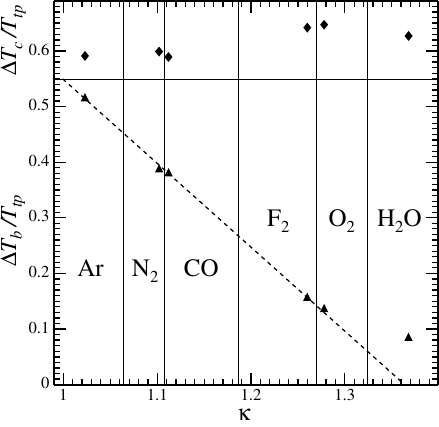}
\caption{The relative temperature drop by which a liquid can be supercooled vs. the parameter $\kappa$ given by (\ref{4.5}), for various fluids. The isoChoric and isoBaric results are marked with the subscripts $_{c}$ and $_{b}$ respectively.}
\label{fig5}
\end{figure}

Fig. \ref{fig5} shows the dependence of $\Delta T/T_{tp}$ on $\kappa$. There
seems to be no evident pattern for isochoric cooling (see the upper part of
Fig. \ref{fig5})---but for the isobaric one, all fluids except water fall onto
the same straight line,%
\[
\frac{\Delta T_{b}}{T_{tp}}=2.06-1.51\,\kappa.
\]
The fact that water is an outlier may reflect its unique physical properties,
or it may arise because the EV model is less accurate for water as it is for
the other fluids examined (as explained in Appendix \ref{appA}).

\section{Comparison with previous work\label{sec5}}

In this section, the results of the present work are compared with those
obtained experimentally and from molecular-dynamics simulations (MDS). Apart
from these, there exists an extensive literature based on classical nucleation
theory (CNT) and similar approaches, but there seems to be no longer a
consensus regarding their application to supercooled liquids. Ref.
\cite{SossoChenCoxFitznerPedevillaZenMichaelides16}, for example, states that
\textquotedblleft at strong supercooling [...] a number of assumptions on
which CNT relies become, if not erroneous, ill-defined\textquotedblright,
which echoes with Ref. \cite{IckesWeltiHooseLohmann15} stating that the CNT
framework \textquotedblleft consists of kinetic and thermodynamic parameters,
of which three are not well-defined\textquotedblright.

The author of the present paper has not worked with CNT and therefore will not
comment on it, except for the difference between its calibration and that of
the EV model. CNT involves \emph{microscopic} parameters, to be determined by
fitting the theoretical nucleation rate to the experimental one---which,
unfortunately, reduces the model's predictive capability. The EV equation, by
contrast, relies on \emph{macroscopic} parameters (the critical and
triple-point characteristics) which can either be easily measured or are
already available in the literature. A model calibrated in this way may not be
as accurate as a data fit, but is accurate enough to reveal interesting trends
underlying supercooling.

In what follows, three qualitative questions will be discussed:

\begin{enumerate}
\item[(A)] According to the EV model, the lowest temperature at which liquid
phase can be observed corresponds to the intersection of the curves labeled
$k=0$ and $k=k_{1.5}$ in Fig. \ref{fig3}. How does this temperature compare
with those at which amorphous states have been observed in other studies?

\item[(B)] The crystallization temperatures computed using the EV model for
three different trajectories are listed in Table \ref{tab2}. The isochoric
value is typically close to the minimum temperature addressed in item (A), but
the isobaric and interfacial values remain to be compared with their
experimental or computational counterparts.

\begin{table}
\begin{ruledtabular}\begin{tabular}{lccccccc}
\rule{0pt}{4.5mm} \vspace{1.5mm} & $T_{s,c}\,(\mathrm{K})$ & $T_{s,b},T_{s,f}\,(\mathrm{K})$ & $T_{tp}\,(\mathrm{K})$\\
\hline
\rule{0pt}{4mm} $\mathrm{Ar}$\vspace{1mm}                 & $\,\,34.2$ & $\,\,40.5$ & $\,\,83.8$\\
\rule{0pt}{4mm} $\mathrm{N}_{2}$\vspace{1mm}              & $\,\,25.4$ & $\,\,38.6$ & $\,\,63.2$\\
\rule{0pt}{4mm} $\mathrm{O}_{2}$\vspace{1mm}              & $\,\,19.2$ & $\,\,46.9$ & $\,\,54.4$\\
\rule{0pt}{4mm} $\mathrm{F}_{2}$\vspace{1mm}              & $\,\,19.1$ & $\,\,45.1$ & $\,\,53.5$\\
\rule{0pt}{4mm} $\mathrm{CO}$\vspace{1mm}                 & $\,\,28.0$ & $\,\,43.1$ & $\,\,68.2$\\
\rule{0pt}{4mm} $\mathrm{H}_{2}\mathrm{O}$\vspace{1mm}    &    $101.9$ &    $249.8$ &    $273.2$
\end{tabular}\end{ruledtabular}
\caption{Spinodal temperatures of various liquids, supercooled along the isochoric, isobaric, and interfacial trajectories (the corresponding temperatures are subscripted $_{s,c}$, $_{s,b}$, and $_{s,f}$, respectively). The last two trajectories are virtually indistinguishable---thus, listed together. For reference, the last column lists the fluids' triple-point temperatures.}
\label{tab2}
\end{table}

\item[(C)] Is there any experimental or computational evidence of the
surface-tension singularity predicted theoretically near the spinodal point?
\end{enumerate}

When comparing the theoretical and experimental results, one should keep in
mind that the former are based on \emph{equilibrium} solutions for \emph{pure}
fluids, whereas the latter may involve \emph{heat and/or mass fluxes}, as well
as \emph{residual gases}.

Note also that $\mathrm{N}_{2}$, $\mathrm{O}_{2}$, $\mathrm{F}_{2}$, and
$\mathrm{CO}$ have so far received little attention in the literature, so the
comparison has to be confined to $\mathrm{Ar}$ and $\mathrm{H}_{2}\mathrm{O}$.

\subsection{Argon\label{sec5.1}}

In what follows, the results of the relevant studies are first summarized and
then commented on in bullet-point form by the author of the present
paper.\medskip

(A) In Ref. \cite{KouchiKuroda90}, a thin film of amorphous argon, together
with argon vapor above it, were placed in a vacuum chamber at $10\,\mathrm{K}%
$, then heated at a rate of $2\,\mathrm{K/min}$ until the film crystallized.
The crystallization began at $20\pm1\,\mathrm{K}$ and ended at $24\pm
1\,\mathrm{K}$. Apart from argon, the vapor contained residual gases:
$\mathrm{H}_{2}\mathrm{O}$ (partial pressure $2.3\times10^{-7}\,\mathrm{Pa}$),
$\mathrm{H}_{2}$ ($1.2\times10^{-7}\,\mathrm{Pa}$), and $\mathrm{N}_{2}$
($0.5\times10^{-7}\,\mathrm{Pa}$). The partial pressure of the argon vapor
increased during crystallization from $0.3\times10^{-7}\,\mathrm{Pa}$ to
$1.0\times10^{-7}\,\mathrm{Pa}$ (both values extracted from Fig. 2 of Ref.
\cite{KouchiKuroda90}).

\begin{itemize}
\item The lowest temperature at which amorphous argon was observed in Ref.
\cite{KouchiKuroda90} is $4\,\mathrm{K}$ below the lowest attainable
temperature for liquid argon predicted by the EV model ($14\,\mathrm{K}%
$).\newline\hspace*{0.5cm}To explain the discrepancy, note that, in Ref.
\cite{KouchiKuroda90}, the pressure of the residual gases exceeded that of
argon by a factor between $10$ (at the beginning of crystallization) and $4$
(at its end). Given the high compressibility of liquid argon---as discussed in
Sect. \ref{sec4.3}---the extra pressure must have made the properties of the
mixture different from those of pure argon---which, in turn, must have altered
the spinodal curve. This alteration explains the discrepancy in the
temperature values mentioned above. In any case, an error of $4\,\mathrm{K}$
may be regarded as acceptable when working in a temperature range located so
far from argon's triple and critical points (which were used to calibrate the
EV model).

\item Note that crystallization in Ref. \cite{KouchiKuroda90} was induced by
\emph{heating} liquid argon, during which its density had to be
\emph{decreasing} (it was not actually measured). This process corresponds to
a trajectory beginning just above the lowest point of the spinodal curve and
moving \emph{upward} and \emph{leftward} until it crosses the boundary of the
no-man's land, triggering crystallization. Note that an isochoric trajectory
originating from the same point would miss the no-man's land and no
crystallization would occur (as evident from Fig. \ref{fig3}).

\item It is unclear whether the heating rate and the resulting evaporation
rate in Ref. \cite{KouchiKuroda90} were small enough for the vapor--liquid
system to be near equilibrium, so the trajectory observed in the experiment
cannot be classified in terms of the present paper (which deals only with
equilibrium states).\medskip
\end{itemize}

(B) In experiments reported in Ref. \cite{MollerSchotteliusCaresanaEtal24},
liquid argon was emitted into vacuum through a nozzle of diameter
$3.5\pm0.5\,\mathrm{\mu m}$, at a temperature $93\pm0.5\,\mathrm{K}$ (i.e.,
approximately $10\,\mathrm{K}$ above the triple point). Due to evaporation,
the temperature of the jet's surface was decreasing with the distance from the
nozzle, and beginning of crystallization was observed at approximately
$68\,\mathrm{K}$ (see Fig. 3 of Ref. \cite{MollerSchotteliusCaresanaEtal24}).

\begin{itemize}
\item Since crystallization in this experiment was attained through cooling,
the corresponding trajectory on the $(n,T)$ plane moves downward and
rightward, eventually crossing the spinodal curve corresponding to the
finite-scale instability. In Fig. \ref{fig3}, this curve is labelled
$k=k_{1.5}$, but the theoretical trajectories crossing it in this figure
cannot be compared to the experimental one, because the jet in the experiment
of Ref. \cite{MollerSchotteliusCaresanaEtal24} was far from equilibrium due to
the presence of strong fluxes.\newline\hspace*{0.5cm}Specifically, there was a
heat flux caused by a significant temperature difference between the jet's
surface ($68\,\mathrm{K}$) and its core ($92\,\mathrm{K}$). This difference
was caused by evaporation, which also caused a significant mass flux through
the liquid--vapor interface. It is safe to assume that the nonequilibrium
processes triggered instability at a temperature well above the theoretical
(interfacial) value of $T_{s,f}=40.5\,\mathrm{K}$ (see Table \ref{tab2}).

\item In the present paper, $T_{s,f}$ was computed for a \emph{flat}
interface, whereas the jet's surface in Ref.
\cite{MollerSchotteliusCaresanaEtal24} was strongly \emph{curved}, which
further impedes quantitative comparison.
\end{itemize}

(C) Unfortunately, no measurements or computations seem to have been reported
for the surface tension of supercooled argon. Furthermore, the surface tension
singularity would generally be difficult to observe in an MDS, as it would
need to simultaneously resolve short scales (the interface and oscillations
near it) and long scales (the region occupied by the oscillations).

\subsection{Water\label{sec5.2}}

(A) As reported in Ref. \cite{JohariHallbruckerMayer87} and numerous other
studies, glassy water can be observed at temperatures as low as
$136\,\mathrm{K}$. A related issue---whether \emph{all} glassy states are
interlinked with a liquid state by a continuous trajectory without
crystallization---has been debated in Refs.
\cite{YueAngell04,KohlBachmannMayerHallbruckerLoerting05,YueAngell05,BachlerGiebelmannAmannwinkelLoerting22}%
, with the latest of these papers concluding that this is indeed possible.

\begin{itemize}
\item The lowest temperature of amorphous water computed in the present paper
is approximately $100\,\mathrm{K}$, which is noticeably lower than the
$136\,\mathrm{K}$ reported in Ref. \cite{JohariHallbruckerMayer87}. The
primary source of this discrepancy is likely the unsuitability of the
nonrotational EV model for water, but one should also note that Ref.
\cite{JohariHallbruckerMayer87} did not specifically search for the glassy
state with the \emph{lowest} temperature.\newline\hspace*{0.5cm}More
generally, this parameter is difficult to find experimentally, as the pressure
at which this state occurs is not known a priori.

\item From the viewpoint of the nonrotational EV model, the fact that any two
amorphous states can be interlinked is evident from Figs. \ref{fig3}%
--\ref{fig4}. It is unclear whether this would change if a model including
molecular rotation were used.\medskip
\end{itemize}

(B) Numerous experiments and MDS studied ice nucleation in supercooled water,
and the highest temperature at which non-zero nucleation rate is observed
corresponds to the spinodal temperature computed in the present work.
Examples, chosen more or less at random, include:

Refs. \cite{RasmussenMackenzieAngellTucker73,AtkinsonMurrayOsullivan16}
experimented with \emph{drops} and placed the cutoff of nucleation at
$235\,\mathrm{K}$ and $237\,\mathrm{K}$, respectively;

Refs. \cite{LiDonadioRussoGalli11,XuPetrikSmithKayKimmel17} reported
simulations and experiments with liquid \emph{films}, obtaining
$243\,\mathrm{K}$ and $238\,\mathrm{K}$, respectively.

\begin{itemize}
\item All of the above experimental and computational temperatures are
slightly lower than the $250\,\mathrm{K}$ calculated in the present paper (see
$T_{s,b}$ and $T_{s.f}$ listed for water in Table \ref{tab2}). The discrepancy
is unlikely to have any cause other than the inaccuracy of the nonrotational
EV model when applied to water.\medskip
\end{itemize}

(C) Refs.
\cite{LueWei06,RogersLeongWang16,WangBinderChenKoopPoschlSuCheng19,MalekPooleSaikavoivod19}
and \cite{VinsHyklHrubyBlahutCelnyCenskyProkopova20} reported computations and
measurements of the surface tension $\gamma$ of the liquid--vapor interface in
supercooled water. In all cases, a considerable increase of $\gamma$ was
reported at low temperatures, well above the predictions of the standard IAPWS
formulation \cite{IAPWS14} extrapolated into the supercooling range. Some of
these papers also report evidence of the so-called second inflection point in
the dependence $\gamma(T)$ located near $T=245\,\mathrm{K}$ (in addition to
the first inflection point at $T=529\,\mathrm{K}$).

\begin{itemize}
\item The above results are consistent with a singularity located below the
second inflection point, but do not prove the singularity.
\end{itemize}

\section{Concluding remarks\label{sec6}}

The EV model used in this paper can be made much more accurate if it is
generalized for \emph{spherically asymmetric, rotating} molecules. There seems
to be no fundamental reason why this cannot be done, and the availability of
this tool is only a matter of time. Such a model should be capable of
quantitatively describing water and other polyamorphic fluids.

An EV model for \emph{multicomponent fluids}, in turn, has already been
developed in Ref. \cite{GrmelaGarciacolin80,GrmelaGarciacolin80b}. It can be
used for the argon-plus-residudals mixture arising in Ref.
\cite{KouchiKuroda90}).

Note, however, that all EV models involve high-dimension multiple integrals
impeding computation. This may be viewed as an argument in favor of more
phenomenological, but simpler, kinetic equations proposed in Refs.
\cite{TakataMatsumotoHiraharaHattori18,TakataMatsumotoHattori21,BhattacharjeeStruchtrupRana24,BusuiocSofonea25,LiGibelliZhang25,TakataTakahashi25}%
. The last of these even describes crystallization and produces `stability
portraits' similar to Figs. \ref{fig3}--\ref{fig4} of the present paper (as
verified by the present author). It would also be interesting to see if
crystallization is described by the moment equations derived in Ref.
\cite{StruchtrupFrezzotti19,StruchtrupFrezzotti22}.

Caution is required, however, when using the diffuse-interface model and
similar hydrodynamic tools: even though they can be rigorously derived from
the EV kinetic model \cite{Giovangigli21}, they perform poorly at and below
near-triple-point temperatures
\cite{BarbanteFrezzottiGibelli14,BarbanteFrezzotti17,Benilov24a}.

Finally, note that liquid--vapor interfaces may be unstable with respect to
disturbances associated with the density gradient and localized in the
interfacial region. In principle, this instability may occur at a temperature
above the spinodal point of the homogeneous liquid and, thus, requires further investigation.

\appendix{}

\section{Calibration of the EV model: the details\label{appA}}

\subsection{Calculating $a$\label{appA.1}}

Observe that expression (\ref{2.28}) for the internal energy is linear in
density, so that the van der Waals constant $a$ can be determined through a
linear fit to the empirical dependence of $e$ vs. $n$.

\begin{figure*}
\includegraphics[width=\columnwidth]{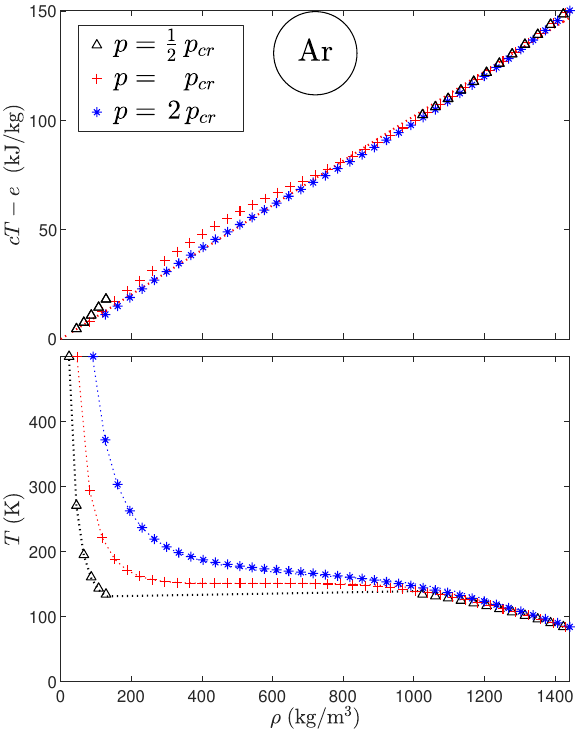}
\includegraphics[width=\columnwidth]{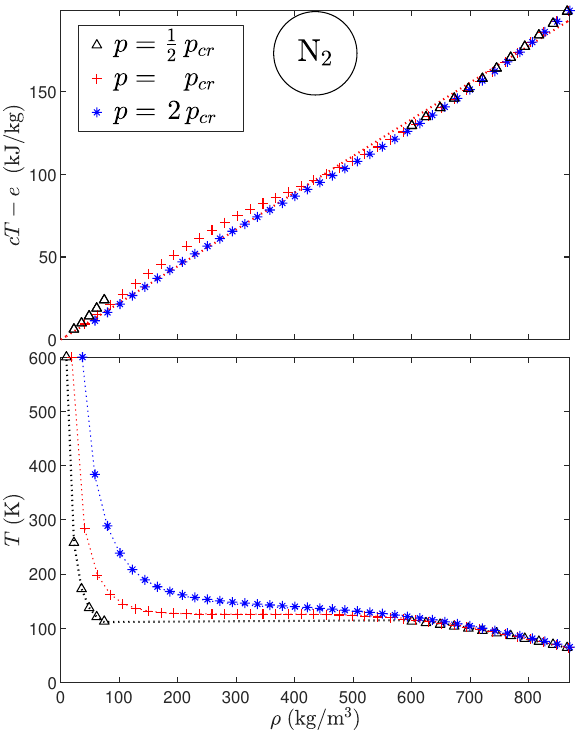}
\caption{The internal energy (upper panels) and the equation of state (lower panels) for argon (left panels) and nitrogen (right panels). The non-connected symbols show the empiric data from Ref. \cite{LindstromMallard97} presented in isobaric form, for three pressure values relative to the critical pressure (see the legends). The gap in the empiric data for $p=p_{cr}/2$ reflects the impossibility/difficulty of measuring parameters of an unstable/metastable fluid. The dotted lines show the curves calculated via the EV model. $N_{A}$ is the Avogadro number.}
\label{fig6}
\end{figure*}

\begin{figure*}
\includegraphics[width=\columnwidth]{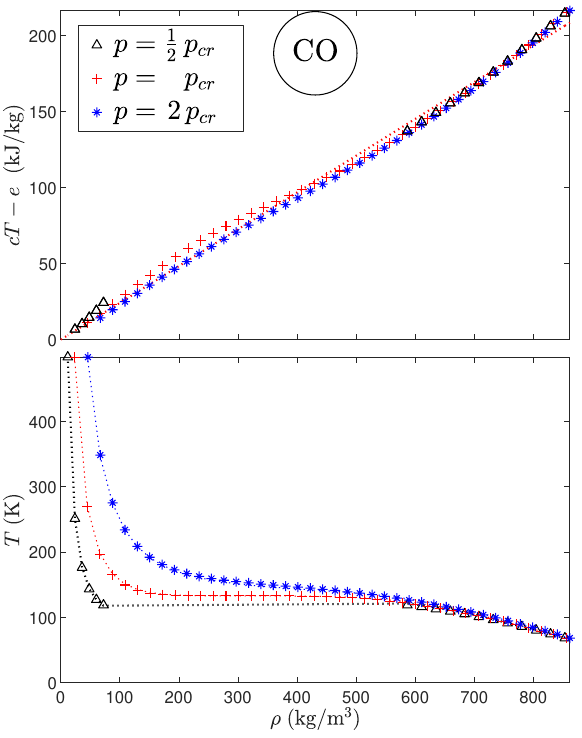}
\includegraphics[width=\columnwidth]{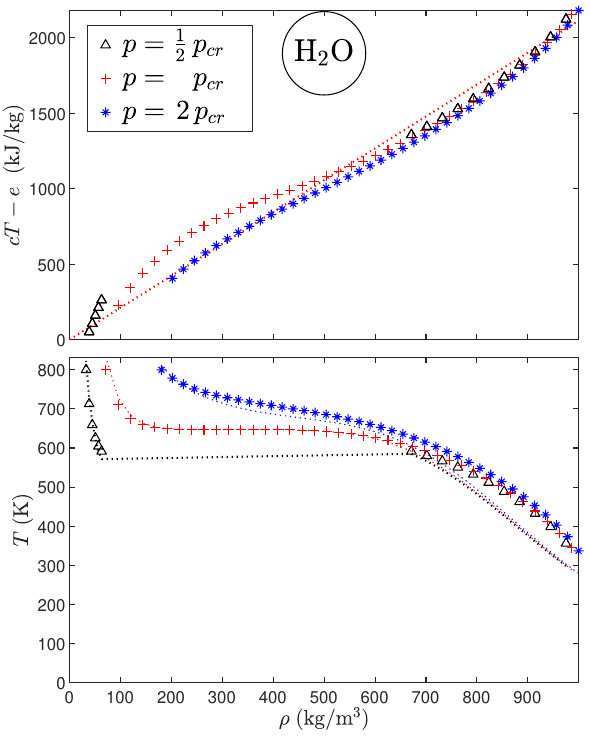}
\caption{The same as in Fig. \ref{fig6}, but for carbon monoxide and water.}
\label{fig7}
\end{figure*}

This has been done for argon, nitrogen, oxygen, fluorine, carbon monoxide, and
water, using, for simplicity, the critical isobar only, $p=p_{cr}$. The
empirical data---in this case and throughout this paper---were taken from Ref.
\cite{LindstromMallard97}. The fitted values of $a$ are listed in Table
\ref{tab1}, and the resulting dependence of $e$ vs. $n$ is plotted, for three
values of $p$ and four different fluids, in the upper panels of Figs.
\ref{fig6}--\ref{fig7}. These plots include the corresponding empirical data,
illustrating that the fitted linear dependence provides a good approximation
over a wide range of pressures and for all fluids considered except water, for
which it is still reasonable. The graphs for $\mathrm{O}_{2}$ an
$\mathrm{F}_{2}$ are not shown, as they are qualitatively similar to that for
$\mathrm{N}_{2}$.

Note that the above calibration procedure assumed the following values of the
heat capacity at constant volume: $C_{V}=3k_{B}/2$ for $\mathrm{Ar}$,
$C_{V}=5k_{B}/2$ for $\mathrm{N}_{2}$, $\mathrm{O}_{2}$, $\mathrm{F}_{2}$, and
$\mathrm{CO}$, and $C_{V}=3k_{B}$ for $\mathrm{H}_{2}\mathrm{O}$. Thus,
although the kinetic model employed in this paper does not account for
molecular rotation, it is calibrated using a heat capacity that does include
rotational degrees of freedom.

Such an approach is strictly consistent only for argon; for other fluids, it
is a device to ensure that the EV model is calibrated to the correct value of
$a$ even in cases where $C_{v}$ is not.

\subsection{Calculating $c_{l}$\label{appA.2}}

With $a$ fixed, one can deduce the coefficients $c_{2}$, $c_{3}$, $c_{4}$, and
$c_{5}$ from the following requirements:

\begin{itemize}
\item the saturated-liquid density $n_{tp}^{(l)}$ computed using
(\ref{2.26})--(\ref{2.27}) for the triple-point temperature $T=T_{tp}$
coincides with its empirical counterpart;

\item the critical-point parameters $T_{cr}$, $n_{cr}$, and $p_{cr}%
$---computed using\begin{widetext}%
\[
\frac{\partial p(n,T)}{\partial n}=0,\qquad\frac{\partial^{2}p(n,T)}{\partial
n^{2}}=0\qquad\text{at}\qquad n=n_{cr},\qquad T=T_{cr},
\]
\end{widetext}with $p(n,T)$ given by (\ref{2.32})---coincide with their
empirical counterparts.
\end{itemize}

\noindent The empirical parameters for the fluids considered, plus their
molecular masses (which also appear in the EV kinetic equation) are listed for
completeness in Table \ref{tab3}. The resulting fitted values of $c_{l}$ are
listed in Table \ref{tab1}.

\begin{table*}
\begin{ruledtabular}\begin{tabular}{lcccccc}
\rule{0pt}{4.5mm} \vspace{1.5mm} & $m~(\mathrm{u})$ & $T_{tp}~(\mathrm{K})$ & $n_{tp}^{(l)}~(\mathrm{mol}/\mathrm{m}^{3})$ & $T_{cr}~(\mathrm{K})$
& $n_{cr} ~(\mathrm{mol}/\mathrm{m}^{3})$ & $p_{cr}~(\mathrm{bar})$\\
\hline
\rule{0pt}{4mm} $\mathrm{Ar}$\vspace{1mm} & $39.9480$ & $\;\;83.81$ & $35.4654$ & $150.69$ & $13.4074$ & $3.39$\\
\rule{0pt}{4mm} $\mathrm{N}_{2}$\vspace{1mm} & $28.0134$ & $\;\;63.15$ & $30.9574$ & $126.19$ & $11.1839$ & $8.71$\\
\rule{0pt}{4mm} $\mathrm{O}_{2}$\vspace{1mm} & $31.9988$ & $\;\;54.36$ & $40.8165$ & $154.58$ & $13.6286$ & $6.53$\\
\rule{0pt}{4mm} $\mathrm{F}_{2}$\vspace{1mm} & $37.9968$ & $\;\;53.48$ & $44.9172$ & $144.41$ & $15.6029$ & $3.72$\\
\rule{0pt}{4mm} $\mathrm{CO}$\vspace{1mm} & $28.0101$ & $\;\;68.16$ & $30.3297$ & $132.86$ & $10.8497$ & $9.71$\\
\rule{0pt}{4mm} $\mathrm{H}_{2}\mathrm{O}$\vspace{1mm} & $18.0153$ & $273.16$ & $55.4969$ & $647.10$ & $17.8737$ & $69.6$
\end{tabular}\end{ruledtabular}
\caption{The parameters (from Ref. \cite{LindstromMallard97}) used for calibration of the EV model.}
\label{tab3}
\end{table*}

Note that, even though the fitted values of $c_{l}$ are derived from the
characteristics of the triple and critical points, the resulting equation of
state agrees well with its empirical counterpart in the whole
triple-to-critical range and for all fluids except water (see the lower panels
of Figs. \ref{fig6}--\ref{fig7}).

\subsection{Calculating $K$\label{appA.3}}

In the simplified versions of the Enskog--Vlasov equation, the
diffuse-interface model \cite{Benilov23a} and Stokes--Vlasov model
\cite{Benilov24a}, the Korteweg constant $K$ was fixed by enforcing the
equality of the theoretical surface tension at the triple point,
$\gamma_{theor}(T_{tp},K)$, to its empirical counterpart, $\gamma
_{empir}(T_{tp})$.

A similar approach is employed in the present paper, but with a modification.
It has turned out that the \emph{exact} equality of $\gamma_{theor}(T_{tp},K)$
and $\gamma_{empir}(T_{tp})$ cannot be enforced, because the former has an
absolute minimum at a certain $K=K_{min}$, and $\gamma(T_{tp},K_{min})$
exceeds $\gamma_{empir}(T_{tp})$. For the fluids examined in this paper except
water, the relative excess is small: $3.1\%$ for $\mathrm{Ar}$, $1.7\%$ for
$\mathrm{N}_{2}$, $3.7\%$ for $\mathrm{O}_{2}$, $4.3\%$ for $\mathrm{F}_{2}$,
and $7.7\%$ for $\mathrm{CO}$---but for $\mathrm{H}_{2}\mathrm{O}$, the
minimum value of $\gamma(T_{tp},K)$ misses its empirical counterpart by a
factor of $2$.

The error of the surface tension of water can only be corrected if the
rotational version of the EV model is used.

\section{Numerical method for boundary-value problem (\ref{3.1})--(\ref{3.6}%
)\label{appB}}

The integrals in expression (\ref{3.2}) for the kernels $F_{l}$ were computed
on a uniform grid inside the cube%
\begin{equation}
\left(  -1<\frac{z_{1}}{D}<1\right)  \times\cdots\times\left(  -1<\frac{z_{l}%
}{D}<1\right)  , \label{C.1}%
\end{equation}
using the Monte Carlo method with the Halton sequence, generated by MATLAB's
haltonset function. The evaluation of $F_{l}$ is the most computationally
expensive part of the calculation.

The solution $n(z)$ of Eq. (\ref{3.1}) was discretized with the same step as
$F_{l}$, within a sufficiently large interval%
\begin{equation}
-Z^{(l)}\leq z\leq Z^{(v)}.\label{C.2}%
\end{equation}
Outside this interval, the values of $n$ were prescribed according to boundary
conditions (\ref{3.4})--(\ref{3.5}),%
\begin{align*}
n(z) &  =n^{(l)}\qquad\text{if}\qquad z<-Z^{(l)},\\
n(z) &  =n^{(v)}\qquad\text{if}\qquad z>Z^{(v)}.
\end{align*}
The integrals on the left-hand side of Eq. (\ref{3.1}) were evaluated using
the rectangle rule, which is \emph{second}-order accurate because the kernels
$F_{l}$ vanish at the boundaries of cube (\ref{C.1}). Enforcing equality
between the left- and right-hand sides of Eq. (\ref{3.1}) at each grid point
within interval (\ref{C.2}) yielded a set of algebraic equations for the grid
values of $n(z)$. The constant $\operatorname{const}$ on the right-hand side
of (\ref{3.1}) was treated as an extra unknown, while condition (\ref{3.6})
was treated as an extra equation.

The resulting system of equations was solved using MATLAB's fsolve function.

\FloatBarrier

\bibliography{../../bib/refs}
% \bibliography{}

\end{document}